\documentclass{appolb}
\usepackage{epsfig}

\usepackage{wrapfig}   %KG
\begin{document}
%\date{\today}
\pagestyle{plain}
%% uncomment the following line to get equations numbered by (sec.num)
%\eqsec
\newcount\eLiNe\eLiNe=\inputlineno\advance\eLiNe by -1
\title{NA49 AND NA61/SHINE EXPERIMENTS: \\ RESULTS AND 
PERSPECTIVES 
%\thanks{Send any remarks to {\tt acta@jetta.if.uj.edu.pl}}%
}
\author{Katarzyna GREBIESZKOW \\{\small for the NA49 and the NA61 
Collaborations}
\address{Faculty of Physics, Warsaw University of Technology,
Koszykowa 75, 00-662~Warsaw, Poland}}
\maketitle

\begin{abstract}

The Super Proton Synchrotron (SPS) covers one of the most
interesting regions of the QCD phase diagram $(T - 
\mu_B)$. On the one
hand there are indications that the energy threshold for 
deconfinement
is reached already at low SPS energies. On the other hand 
theoretical
calculations locate the QCD critical end-point at energies accessible
at the SPS. In this paper the NA49 signatures of the onset of
deconfinement are presented. Results are shown on pion production, the
kaon to pion ratio, slopes of transverse mass spectra ("temperature"),
and event-by-event particle ratio fluctuations versus collision energy,
for central $Pb+Pb$ interactions. Next we show possible indications of
the critical point in event-by-event mean transverse momentum and
multiplicity fluctuations. Finally we discuss the future ion program of
the NA61/SHINE experiment (energy scan with light ions).

\end{abstract}

\clearpage

\vspace{0.5cm}
\noindent
{\bf The NA49 Collaboration:} \\
T.~Anticic$^{23}$, B.~Baatar$^{8}$,D.~Barna$^{4}$,
J.~Bartke$^{6}$, L.~Betev$^{10}$, H.~Bia{\l}\-kowska$^{20}$,
C.~Blume$^{9}$,  B.~Boimska$^{20}$, M.~Botje$^{1}$,
J.~Bracinik$^{3}$, P.~Bun\v{c}i\'{c}$^{10}$,
V.~Cerny$^{3}$, P.~Christakoglou$^{2}$,
P.~Chung$^{19}$, O.~Chvala$^{14}$,
J.G.~Cramer$^{16}$, P.~Csat\'{o}$^{4}$, P.~Dinkelaker$^{9}$,
V.~Eckardt$^{13}$,
%H.G.~Fischer$^{10}$,
Z.~Fodor$^{4}$, P.~Foka$^{7}$,
V.~Friese$^{7}$, J.~G\'{a}l$^{4}$,
M.~Ga\'zdzicki$^{9,11}$, V.~Genchev$^{18}$,
E.~G{\l}adysz$^{6}$, K.~Grebieszkow$^{22}$,
S.~Hegyi$^{4}$, C.~H\"{o}hne$^{7}$,
K.~Kadija$^{23}$, A.~Karev$^{13}$, D.~Kikola$^{22}$,
V.I.~Kolesnikov$^{8}$, E.~Kornas$^{6}$,
R.~Korus$^{11}$, M.~Kowalski$^{6}$,
M.~Kreps$^{3}$, A.~Laszlo$^{4}$,
R.~Lacey$^{19}$, M.~van~Leeuwen$^{1}$,
P.~L\'{e}vai$^{4}$, L.~Litov$^{17}$, B.~Lungwitz$^{9}$,
M.~Mackowiak$^{22}$
M.~Makariev$^{17}$, \\ A.I.~Malakhov$^{8}$,
M.~Mateev$^{17}$, G.L.~Melkumov$^{8}$, A.~Mischke$^{1}$,
M.~Mitrovski$^{9}$,
J.~Moln\'{a}r$^{4}$, St.~Mr\'owczy\'nski$^{11}$, V.~Nicolic$^{23}$,
G.~P\'{a}lla$^{4}$, A.D.~Panagiotou$^{2}$, D.~Panayotov$^{17}$,
A.~Petridis$^{2,\ast}$, W.~Peryt$^{22}$, M.~Pikna$^{3}$,
J.~Pluta$^{22}$,
D.~Prindle$^{16}$,
F.~P\"{u}hlhofer$^{12}$, R.~Renfordt$^{9}$,
C.~Roland$^{5}$, G.~Roland$^{5}$,
M. Rybczy\'nski$^{11}$, A.~Rybicki$^{6}$,
A.~Sandoval$^{7}$, N.~Schmitz$^{13}$, T.~Schuster$^{9}$,
P.~Seyboth$^{13}$,
F.~Sikl\'{e}r$^{4}$, B.~Sitar$^{3}$, E.~Skrzypczak$^{21}$,
M.~Slodkowski$^{22}$,
G.~Stefanek$^{11}$, R.~Stock$^{9}$, C.~Strabel$^{9}$,
H.~Str\"{o}bele$^{9}$, T.~Susa$^{23}$,
I.~Szentp\'{e}tery$^{4}$, J.~Sziklai$^{4}$, M.~Szuba$^{22}$,
P.~Szymanski$^{10,20}$,
V.~Trubnikov$^{20}$, M.~Utvic$^{9}$, D.~Varga$^{4,10}$,
M.~Vassiliou$^{2}$, G.I.~Veres$^{4,5}$, G.~Vesztergombi$^{4}$,
%S.~Wenig$^{10}$,
D.~Vrani\'{c}$^{7}$,
Z.~W{\l}odarczyk$^{11}$, A.~Wojtaszek-Szwarc$^{11}$, I.K.~Yoo$^{15}$

\vspace{0.5cm}
\noindent
$^{1}$NIKHEF, Amsterdam, Netherlands. \\
$^{2}$Department of Physics, University of Athens, Athens, Greece.\\
$^{3}$Comenius University, Bratislava, Slovakia.\\
$^{4}$KFKI Research Institute for Particle and Nuclear Physics,
Budapest, Hungary.\\
$^{5}$MIT, Cambridge, USA.\\
$^{6}$Institute of Nuclear Physics, Cracow, Poland.\\
$^{7}$Gesellschaft f\"{u}r Schwerionenforschung (GSI), Darmstadt,
Germany.\\
$^{8}$Joint Institute for Nuclear Research, Dubna, Russia.\\
$^{9}$Fachbereich Physik der Universit\"{a}t, Frankfurt, Germany.\\
$^{10}$CERN, Geneva, Switzerland.\\
$^{11}$Institute of Physics, Jan Kochanowski University, Kielce,
Poland.\\
$^{12}$Fachbereich Physik der Universit\"{a}t, Marburg, Germany.\\
$^{13}$Max-Planck-Institut f\"{u}r Physik, Munich, Germany.\\
$^{14}$Institute of Particle and Nuclear Physics, Charles University,
Prague, Czech Republic.\\
$^{15}$Department of Physics, Pusan National University, Pusan, Republic
of Korea.\\
$^{16}$Nuclear Physics Laboratory, University of Washington, Seattle,
WA, USA.\\
$^{17}$Atomic Physics Department, Sofia University St. Kliment Ohridski,
Sofia, Bulgaria.\\
$^{18}$Institute for Nuclear Research and Nuclear Energy, Sofia,
Bulgaria.\\
$^{19}$Department of Chemistry, Stony Brook Univ. (SUNYSB), Stony Brook,
USA.\\
$^{20}$Institute for Nuclear Studies, Warsaw, Poland.\\
$^{21}$Institute for Experimental Physics, University of Warsaw, Warsaw,
Poland.\\
$^{22}$Faculty of Physics, Warsaw University of Technology, Warsaw,
Poland.\\
$^{23}$Rudjer Boskovic Institute, Zagreb, Croatia.\\
$^{\ast}$deceased

\vspace{0.8cm}
\noindent
{\bf The NA61/SHINE Collaboration:} \\
N.~Abgrall${}^{22}$,
A.~Aduszkiewicz${}^{23}$,
B.~Andrieu${}^{11}$,
T.~Anticic${}^{13}$,
N.~Antoniou${}^{18}$,
J.~Argyriades${}^{22}$,
A.~G.~Asryan${}^{15}$,
B.~Baatar${}^{9}$,
A.~Blondel${}^{22}$,
J.~Blumer${}^{5}$
% ...

\vspace{0.5cm}
\noindent
${}^{ 1}$Cape Town University, Cape Town, South Africa \\
${}^{ 2}$ETH, Zurich, Switzerland \\
${}^{ 3}$Fachhochschule Frankfurt, Frankfurt, Germany \\
${}^{ 4}$Faculty of Physics, University of Sofia, Sofia, Bulgaria \\
${}^{ 5}$Forschungszentrum Karlsruhe, Karlsruhe, Germany \\
${}^{ 6}$Institute for Nuclear Research, Moscow, Russia \\
${}^{ 7}$Institute for Particle and Nuclear Studies, KEK, Tsukuba,  
Japan \\
${}^{ 8}$Jagiellonian University, Cracow, Poland  \\
${}^{ 9}$Joint Institute for Nuclear Research, Dubna, Russia \\
${}^{10}$KFKI Research Institute for Particle and Nuclear Physics, 
Budapest, Hungary \\
${}^{11}$LPNHE, University of Paris VI and VII, Paris, France \\
${}^{12}$University of Silesia, Katowice, Poland  \\
${}^{13}$Rudjer Boskovic Institute, Zagreb, Croatia \\
${}^{14}$Soltan Institute for Nuclear Studies, Warsaw, Poland \\
${}^{15}$St. Petersburg State University, St. Petersburg, Russia \\
${}^{16}$State University of New York, Stony Brook, USA \\
${}^{17}$Jan Kochanowski University in  Kielce, Poland \\
${}^{18}$University of Athens, Athens, Greece \\
${}^{19}$University of Bergen, Bergen, Norway \\
${}^{20}$University of Bern, Bern, Switzerland \\
${}^{21}$University of Frankfurt, Frankfurt, Germany \\
${}^{22}$University of Geneva, Geneva, Switzerland \\
${}^{23}$University of Warsaw, Warsaw, Poland \\
${}^{24}$Warsaw University of Technology, Warsaw, Poland  \\

\section{Introduction}

It is a well established fact that matter exists in different states. 
For strongly interacting matter at least three are expected: 
normal nuclear matter (liquid), hadron gas (HG), and a system of 
deconfined quarks and gluons (eventually the quark-gluon plasma). One of 
the most important goals of high-energy heavy-ion collisions is to 
establish the phase diagram of strongly interacting matter by finding 
the possible phase boundaries and critical points. In principle, we want 
to produce the quark-gluon plasma (QGP) and analyze its properties and 
the transition between QGP and HG.

\begin{wrapfigure}{r}{7.cm}
\vspace{-0.2cm}
\includegraphics[scale=0.35]{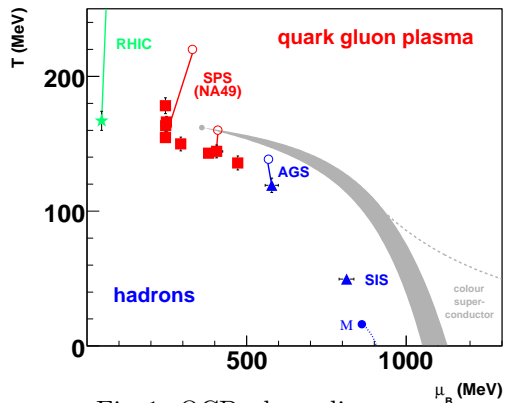}
\vspace{-1.0cm}
\caption[]{QCD phase diagram.}
\label{phas_full}
\end{wrapfigure}

The phase diagram of strongly interacting matter (Fig.~\ref{phas_full}) 
is most often presented in terms of temperature $T$ and baryochemical
potential $\mu_B$. It is believed that for large values of
$\mu_B$ the phase transition is of the first order (gray band) and for 
low $\mu_B$ values a rapid but continuous transition (cross-over). A 
critical point of second order (CP) separates those two regions. The open
points in Fig.~\ref{phas_full} are hypothetical locations 
reached by the high-density matter droplet after dissipation of the 
energy of the incident nucleons from where the evolution of the expanding
and cooling fireball starts. The closed symbols are chemical freezout points 
\cite{beccatini} (chemical composition fixed, the end of inelastic 
interactions). The fireball then expands further until
thermal (kinetic) freeze-out takes place (particle momenta are fixed).

The Super Proton Synchrotron (SPS) covers one of the most
interesting regions of the QCD phase diagram $(T - \mu_B)$. On the one
hand it is expected that the energy threshold for deconfinement is
reached already at low SPS energies (see the open point hitting the 
transition line in Fig.~\ref{phas_full}). On the other hand  
lattice QCD calculations locate the critical point of strongly 
interacting matter in the SPS energy range \cite{fodor_latt_2004}.

\section{NA49 experiment}

NA49 \cite{na49_nim}, operating in 1994-2002, was one of the fixed 
target heavy-ion experiments at the CERN SPS. The main components of 
the detector were four large volume time projection chambers (TPC). The 
Vertex TPCs (VTPC-1 and VTPC-2), were located in the magnetic field of 
two super-conducting dipole magnets. Two other TPCs (MTPC-L and MTPC-R) 
were positioned downstream of the magnets symmetrically to the beam 
line. The NA49 TPCs allowed precise measurements of particle momenta 
$p$ with a resolution of $\sigma(p)/p^2 \cong (0.3-7)\cdot10^{-4}$ 
(GeV/c)$^{-1}$. A precise measurement of specific energy loss ($dE/dx$) 
in the region of relativistic rise was possible in the TPCs. Time of 
Flight (TOF) walls supplemented particle identification close to 
mid-rapidity. The typical $dE/dx$ resolution was $\sigma(dE/dx) / 
\langle dE/dx \rangle \approx 0.04$, the typical ToF resolution $\sigma 
(ToF) \approx 60$~ps, and invariant mass resolution (for identification 
of particles via decay topology) was $\sigma (m_{inv}) \approx 5$ MeV.  
The centrality of nuclear collisions was selected via measurement of
the energy of the projectile spectator nucleons in the Forward
Calorimeter. The NA49 acceptance covers the forward hemisphere, but because of 
the symmetry of nucleus+nucleus ($A+A$) collisions this nevertheless allows to 
extract $4 \pi$ integrated multiplicities.

\section{NA49 evidence for the Onset of Deconfinement}

\subsection{Kink, horn, step and SMES}

In 2002 the NA49 experiment completed the SPS energy scan of   
central $Pb+Pb$ collisions. This program was originally motivated by 
predictions of the Statistical Model of the Early Stage (SMES) 
\cite{mg_model} assuming that the energy threshold for 
deconfinement (the lowest energy sufficient to create a partonic 
system) is located at low SPS energies. Several structures were 
expected within SMES: the kink in pion production (due to increased 
entropy production), the horn in the strangeness to entropy ratio, and the step 
in the inverse slope parameter of transverse momentum spectra 
(constant temperature and pressure in a mixed phase).

\begin{wrapfigure}{r}{7.cm}
\vspace{-0.5cm}
\includegraphics[scale=0.35]{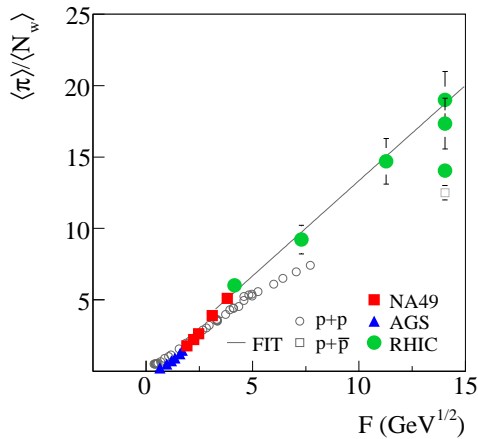}
\vspace{-1.1cm}
\caption[]{Energy ($F$) dependence of the mean pion multiplicity per 
wounded nucleon \cite{kpi_paper} in full phase space ($4\pi$).}
\label{kink}
\end{wrapfigure}

Figure~\ref{kink} shows production of charged pions (the total entropy is 
carried mainly by pions \footnote{In SMES the total entropy and the total 
strangeness are the same before and after hadronization (the entropy cannot 
decrease during the transition from QGP to hadron gas), therefore pions 
measure the early stage entropy.}) $\langle \pi \rangle = 1.5 (\langle \pi^{+} 
\rangle + \langle \pi^{-} \rangle) $ normalized to the number of 
wounded nucleons versus Fermi variable $F$ ($F \approx 
(s_{NN})^{1/4}$). In SMES, this ratio is proportional to the 
effective number of degrees of freedom ($NDF$) to the power of 1/4.
For central $A+A$ collisions ($Pb+Pb$ for SPS or $Au+Au$ for AGS and RHIC) a 
change of slope around 30$A$~GeV is visible (slope in $A+A$ increases from 
$\approx$ 1 (AGS) to $\approx$ 1.3 (top SPS+RHIC) - consistent with 
increase by a factor of 3 in $NDF$). Such an increase is not observed for
$p + p (\bar{p})$ reactions. The increase in $NDF$, when going from hadron 
gas to QGP, may be interpreted as a consequence of the activation of partonic 
degrees of freedom.

\begin{figure}[ht]
\centering
\vspace{-0.3cm}
\includegraphics[width=0.49\textwidth]{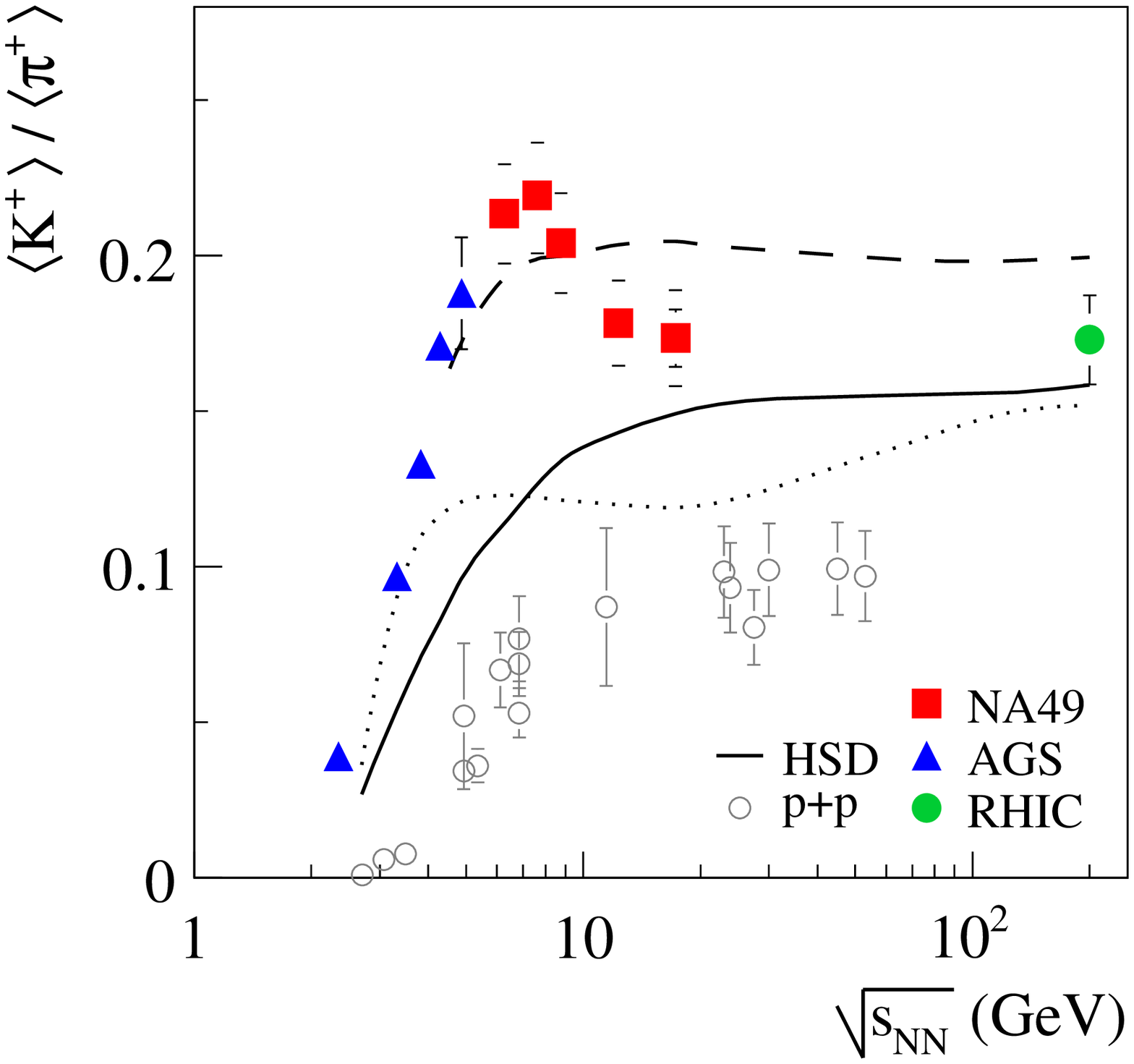}
\includegraphics[width=0.49\textwidth]{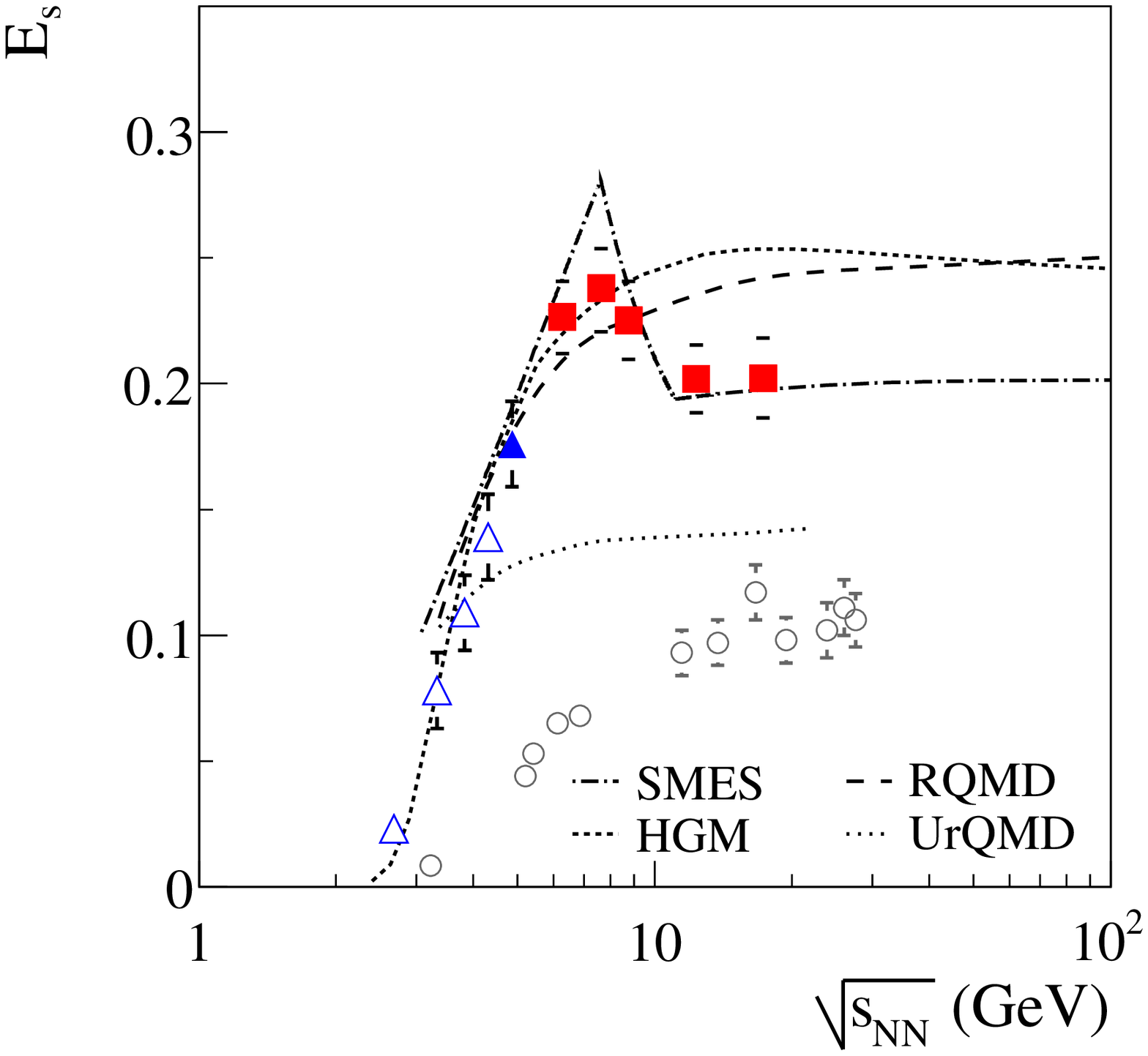}
\vspace{-0.5cm}
\caption[]{Left: Energy dependence of the $\langle K^{+} \rangle /
\langle \pi^{+} \rangle $ ratio in full phase space ($4\pi$). Right:
$E_s$ measure of strangeness to entropy ratio versus energy in $4\pi$ 
\cite{kpi_paper}.}
\label{horn}
\end{figure}

Figure~\ref{horn} (left) presents the $\langle K^{+} \rangle / \langle \pi^{+} 
\rangle $ ratio in full phase space ($4\pi$) versus energy. In 
SMES, the ratio is proportional to strangeness/entropy densities. Results for 
$A+A$ are very different from the results for $p+p$ and show a sharp 
peak in $\langle K^{+} \rangle / \langle \pi^{+} \rangle $ at 30$A$ GeV. 
This peak is even more pronounced at mid-rapidity (see 
\cite{kpi_paper}). Recently, these intriguing NA49 results on pion 
and kaon yields (at mid-rapidity) were confirmed at 
$\sqrt{s_{NN}}$ = 9.2 GeV and 19.6 GeV by the STAR 
experiment \cite{star_lowen}. The string hadronic models 
(curves in Fig.~\ref{horn} (left)) do not reproduce the data. 
The hadron gas model (HGM) \cite{HGM_1} using a parameterization
of the energy dependence $T_{ch}$ and $\mu_B$ based on fits to hadron yields
produces a broad  maximum in the $\langle K^{+} \rangle / \langle 
\pi^{+} \rangle $ ratio as a consequence of saturating $T_{ch}$ and 
decreasing $\mu_B$ with increasing energy (limiting temperature reached 
somewhere at the SPS). However, this version of the model overestimates the 
relative kaon yields from 30$A$ GeV on. The latest extension of the HGM 
\cite{HGM_2} yields and improves fit of the energy dependence $\langle 
K^{+} \rangle / \langle \pi^{+} \rangle $ by inclusion of controversial 
\cite{PDG2008} $\sigma$ state and {\it unmeasured} 
resonances above 3 GeV in the model hadron spectrum  (see 
conference slides or \cite{HGM_2}). 

The measure which much better reflects the total strangeness to entropy 
ratio in the SPS energy range is 
$E_s = (\langle K \rangle + \langle \Lambda \rangle)/ \langle 
\pi \rangle$, proposed in Ref.~\cite{mg_model}, and calculated from $\pi$, 
$K$, and $\Lambda$ yields in $4\pi$ acceptance. The $E_s$ ratio can be 
directly and quantitatively compared to SMES predictions. 
Figure~\ref{horn} (right) shows a distinct peak in $E_s$ at 30$A$ GeV. This
behavior is described (predicted) only by the model assuming a phase 
transition (i.e. SMES), where the maximum, called 'horn', is the result of 
the decrease of strangeness carrier masses 
in the QGP ($m_s < m_{\Lambda, K, ...}$) and the change in the number of 
degrees of freedom when reaching the deconfined state.

\begin{figure}[ht]
\centering
\vspace{-0.3cm}
\includegraphics[width=0.49\textwidth]{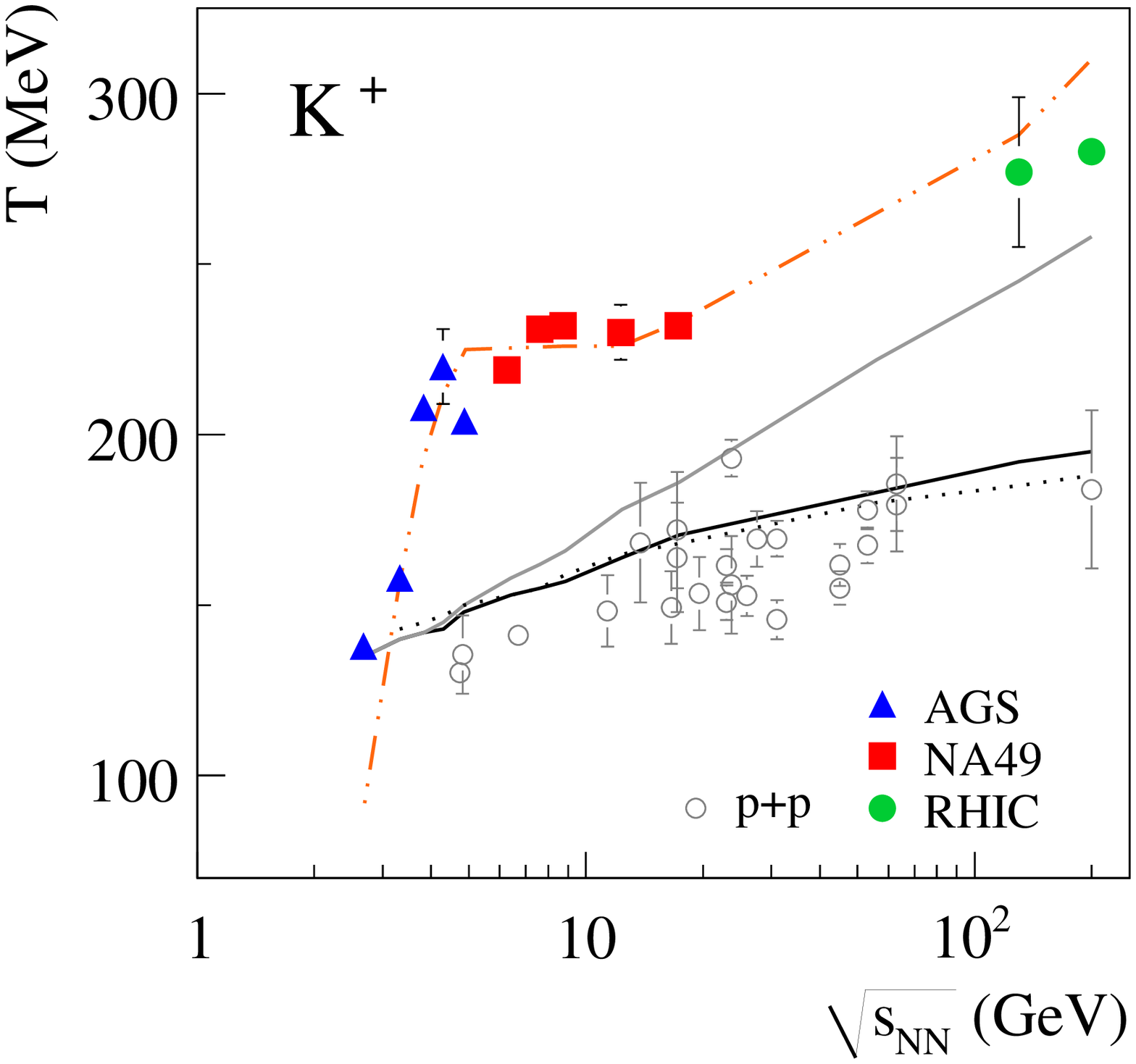}
\includegraphics[width=0.49\textwidth]{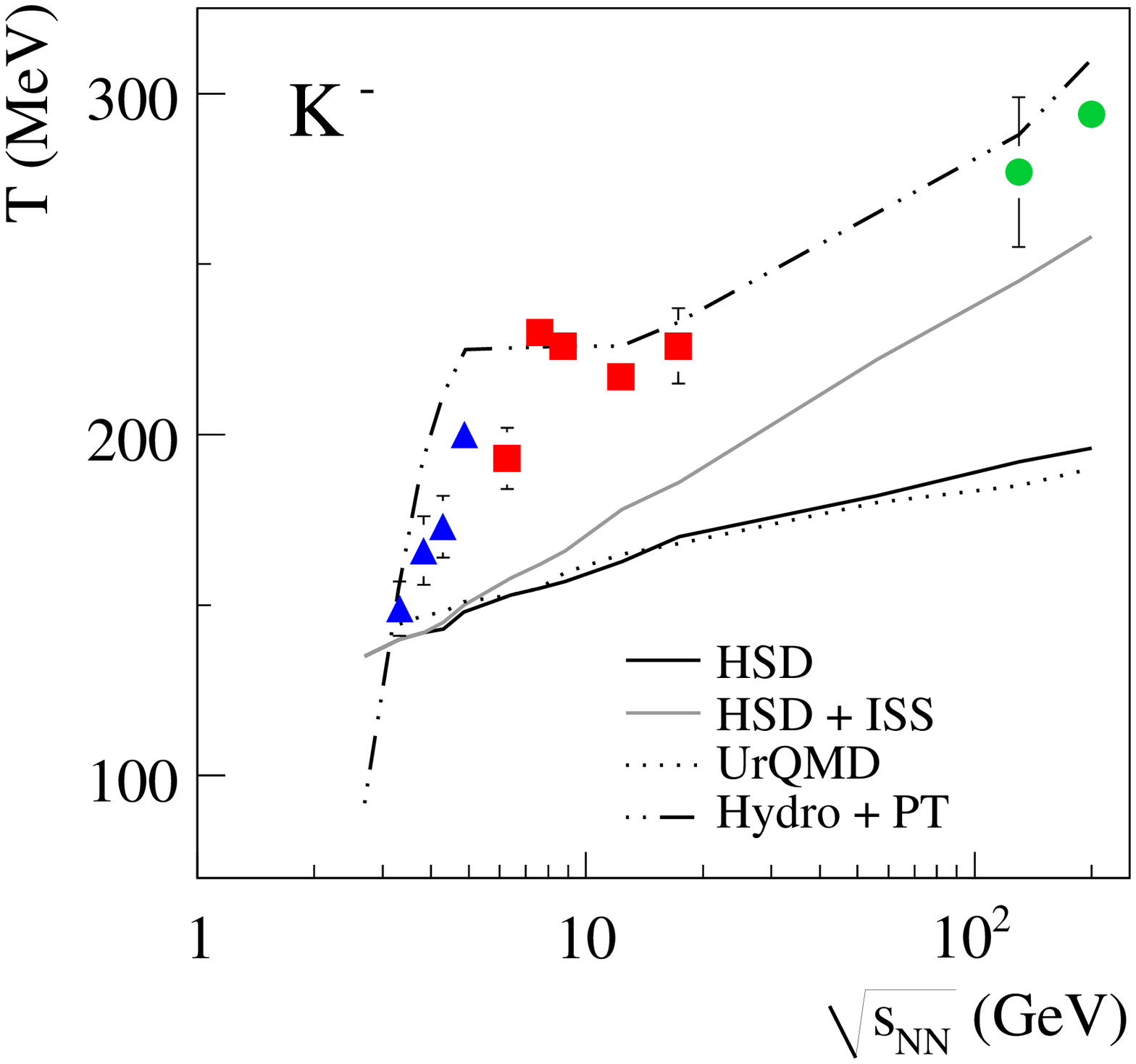}
\vspace{-0.5cm}
\caption[]{Energy dependence of the inverse slope parameter $T$ of the
transverse mass spectra of $K^{+}$ and $K^{-}$ mesons \cite{kpi_paper}.}
\label{step}
\end{figure}

Figure~\ref{step} presents inverse slope parameters ($T$) of transverse 
mass spectra~\footnote{Transverse mass spectra were parametrized by 
$dn/(m_T dm_T) 
= C \cdot exp (-m_T / T)$; fits were done close to mid-rapidity.} 
of positively and negatively charged kaons. For $A+A$ data one can see a 
strong rise at AGS, plateau at SPS, and rise towards RHIC energies. Such 
structure is not observed for $p+p$ collisions. The 
plateau is consistent with constant temperature and pressure in the mixed 
phase (latent heat) \cite{smes_temp}. In fact, this structure 
strongly resembles the behavior of water, where a plateau can be observed 
in the temperature when heat is added. Models without phase transition do 
not reproduce the $A+A$ data, but a hydrodynamical model 
incorporating a deconfinement phase transition at SPS energies 
\cite{smes_hydro} describes the results in Fig.~\ref{step} quite well. 
The step-like feature is also present in the energy dependence of 
$m_T - m$ of protons and pions (see conference 
slides or \cite{kpi_paper}).

\subsection{Event-by-event particle ratio fluctuations}

Hadron ratios characterize the chemical composition of the fireball and 
are not affected by hadronic re-interaction when looking at conserved 
quantities such as baryon number, strangeness. We expect a change of 
particle (e.g. strangeness) production properties close to the phase 
transition. In principle, two distinct event classes (with/without QGP) 
or the mixed phase (coexistence region of hadronic and partonic matter for 
$1^{st}$ order phase transition) may be reflected in larger 
event-by-event fluctuations.

\begin{figure}[ht]
\centering
\vspace{-0.1cm}
\includegraphics[width=0.49\textwidth]{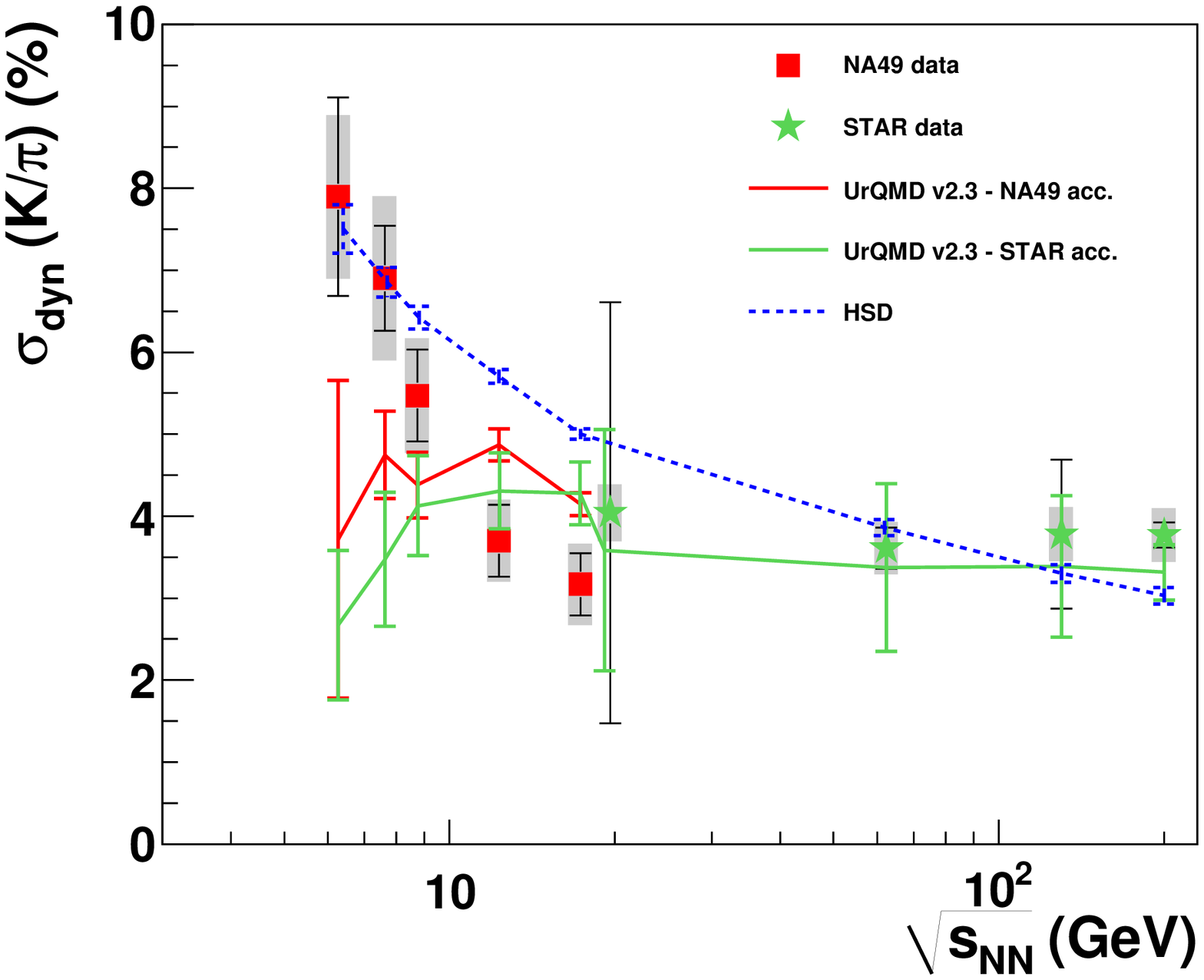}
\includegraphics[width=0.49\textwidth]{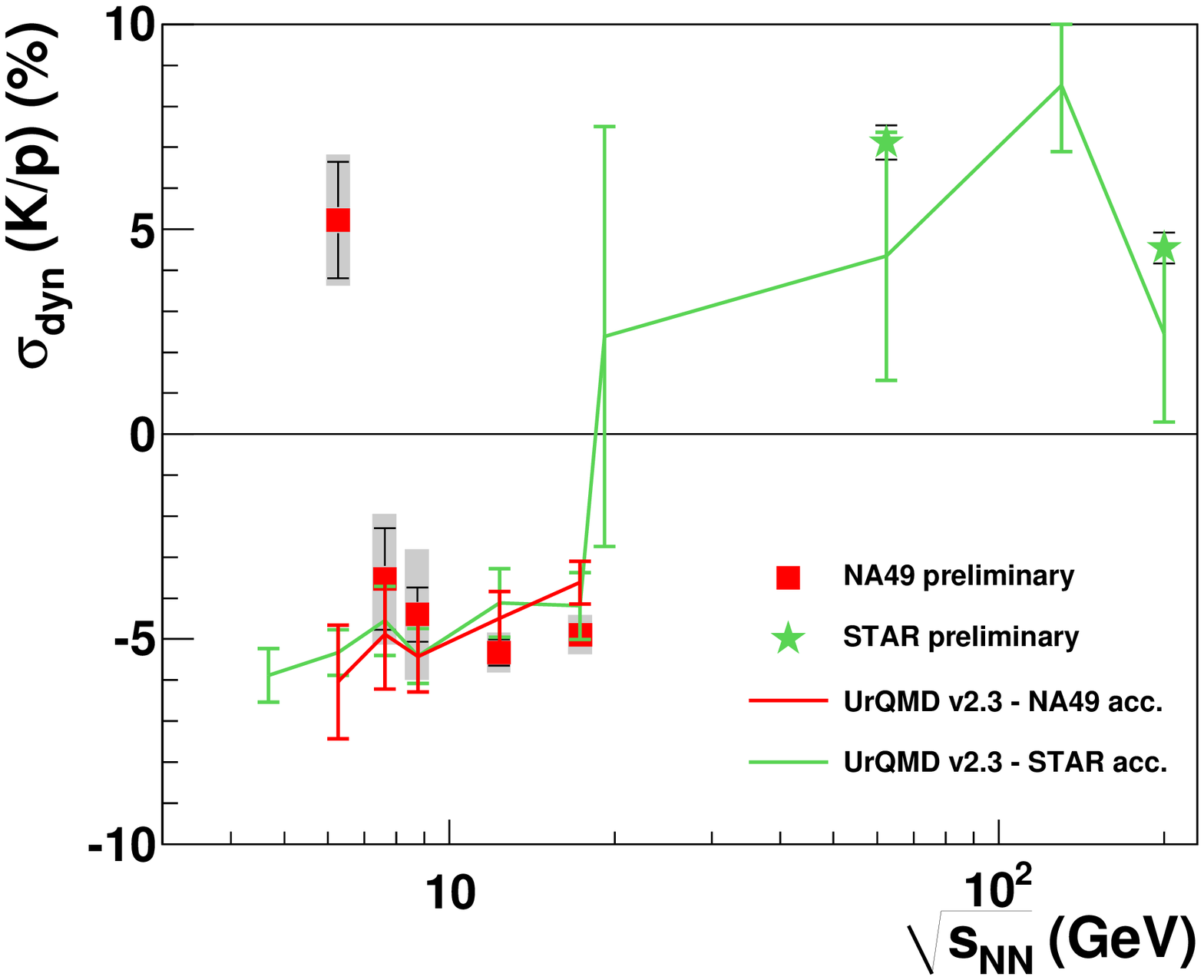}
\vspace{-0.3cm}
\caption[]{Energy dependence of dynamical $K/\pi$ (left) and $K/p$ 
(right) fluctuations for 3.5\% most central $Pb+Pb$ (NA49) 
and for $Au+Au$ interactions (STAR) \cite{kpi_fluct}.}
\label{fluct_ratio}
\end{figure}

The NA49 experiment used $\sigma_{dyn}$ \cite{kpi_fluct} to quantify
event-by-event particle ratio fluctuations. $\sigma_{dyn}$ measures the 
difference between widths of particle ratio distributions for data and 
for artificially produced mixed events, where only statistical 
fluctuations are present. Figure~\ref{fluct_ratio} (left) shows that the
dynamical fluctuations in the $K/\pi$ ratio are positive. One observes
a steep rise towards low SPS energies but no significant change
from top SPS to RHIC energies. The rise towards low energies is not 
reproduced by the UrQMD transport model (there is no significant 
acceptance dependence). The HSD transport model catches the trend but 
overpredicts the data at high SPS energies. Dynamical fluctuations of 
the $K/p$ ratio (Fig.~\ref{fluct_ratio} (right)) exhibit two 
sign changes: a positive plateau at RHIC energies changes to a negative plateau at 
higher SPS energies, followed by a jump to positive value at the lowest SPS energy 
(20$A$ GeV). High energy SPS and RHIC data are reproduced by the UrQMD model, 
however the jump at 20$A$ GeV is not described (this jump between SPS and 
RHIC is not due to acceptance). The values of dynamical event-by-event 
fluctuations of the $p/\pi$ ratio (see conference slides or 
\cite{kpi_fluct}) are negative (both at SPS and at RHIC energies) and 
can be reproduced by the UrQMD model (understood in terms of baryon 
resonance decays) in the SPS energy range.

\section{NA49 indications for the Critical Point}

\begin{wrapfigure}{r}{7.cm}
\centering
\vspace{-0.5cm}
\includegraphics[scale=0.38]{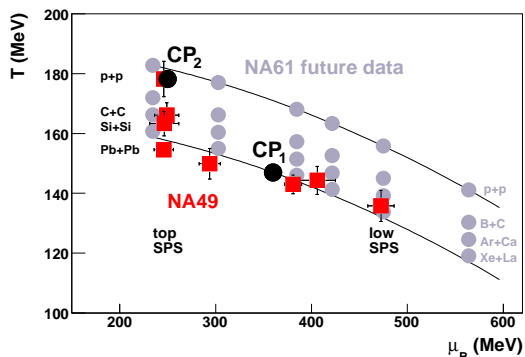}
\vspace{-0.7cm}
\caption[]{Chemical freeze-out points in NA49 (squares) and those
expected in NA61 (circles). See the text for details.}
\label{cp1cp2}
\end{wrapfigure}

Lattice QCD calculations locate the QCD critical point at energies
accessible at the CERN SPS \cite{fodor_latt_2004}. 
In "normal" liquids (including water) the critical point can be
rather easily detected via the critical opalescence phenomenon (scattering
of light on critical long wavelength density fluctuations). Over the
past years several experimental observables were proposed to look
for the CP in heavy ion collisions. Among them are fluctuations
of mean transverse momentum and multiplicity \cite{SRS},
pion pair (sigma mode) intermittency, elliptic flow of baryons and 
mesons, and transverse mass spectra of baryons and anti-baryons
\cite{kg_qm09}. One should also note that for strongly interacting
matter the maximum effect of the CP is expected when the freeze-out happens
near the critical point.

The position of the chemical freeze-out point in the $(T - \mu_B)$ diagram
can be varied by changing the energy and the size of the colliding
system as presented in Fig.~\ref{cp1cp2}. Therefore we analyzed
in NA49 the energy dependence of the proposed CP sensitive
observables for central $Pb+Pb$ collisions (beam energies 20$A$-158$A$
GeV), and their system size dependence ($p+p$, $C+C$, $Si+Si$, and 
$Pb+Pb$) at the highest SPS energy. Figure~\ref{cp1cp2} shows $CP_{1}$ 
and $CP_{2}$ that were considered in NA49 as possible locations of the 
critical point: $CP_1$ with $\mu_B$ from lattice QCD 
calculations \cite{fodor_latt_2004} and $T$ on the empirical freeze-out
line; $CP_2$ as the chemical freeze-out point of $p+p$ reactions 
at 158$A$ GeV (assuming that this freeze-out point may be 
located on the phase transition line).

\subsection{Average $p_T$ and multiplicity fluctuations}

At the critical point enlarged fluctuations of multiplicity and mean transverse
momentum are expected \cite{SRS}. In NA49 we used the $\Phi_{p_T}$
fluctuation measure \cite{fluct_size, fluct_energy} and the scaled
variance $\omega$ of multiplicity distribution \cite{omega_size,
omega_energy} to study $p_T$ and $N$ fluctuations, respectively. For a
system of independently emitted particles (no inter-particle
correlations) $\Phi_{p_T}$ is equal to zero. For a Poisson multiplicity
distribution $\omega$ equals 1.
If $A+A$ reactions are a superposition of independent $N+N$ collisions then 
$\Phi_{p_T}$($A+A$) = $\Phi_{p_T}$ ($N+N$), whereas $\omega$($A+A$) = 
$\omega$($N+N$) + $\langle n \rangle \omega_{part}$, where $\langle n  
\rangle$ is the mean multiplicity of hadrons from a single $N+N$ collisions and 
$\omega_{part}$ represents
fluctuations in $N_{part}$. The above equations suggest that while
$\Phi_{p_T}$ is independent of $N_{part}$ fluctuations, $\omega$ is
strongly dependent on them. In the NA49 fixed target
experiment $N_{part}^{proj}$ can be fixed (spectator energy measured by
the Forward Calorimeter), whereas $N_{part}^{targ}$ cannot be measured.
It was shown \cite{konchak} that fluctuations of $N_{part}^{targ}$ can be suppressed
only by selection of very central collisions. Therefore multiplicity
fluctuations are presented ($\omega$) for very central
(1\%) collisions.

\begin{figure}[ht]
\centering
\vspace{-0.3cm}
\includegraphics[scale=0.6]{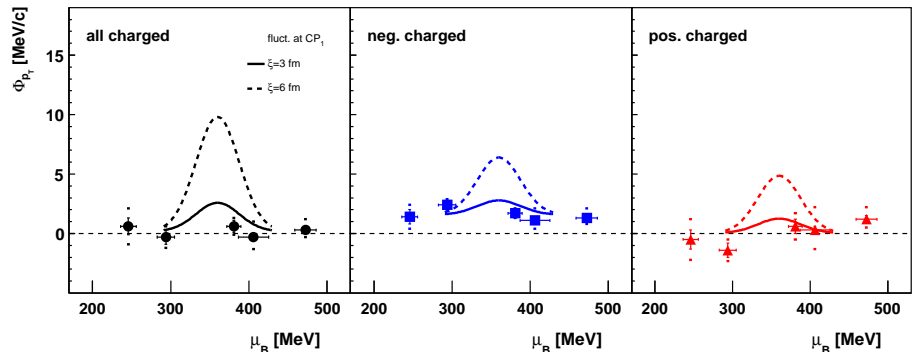}
\vspace{-0.5cm}
\caption[]{Energy dependence of $\Phi_{p_T}$ for 7.2\% most central
$Pb+Pb$ collisions \cite{fluct_energy}. Lines correspond to $CP_1$ predictions 
(see text); their base-lines are the mean $\Phi_{p_T}$ values for 5 energies.}
\label{fiptmb}
\end{figure}

\begin{figure}[ht]
\centering
\vspace{-0.5cm}
\includegraphics[scale=0.6]{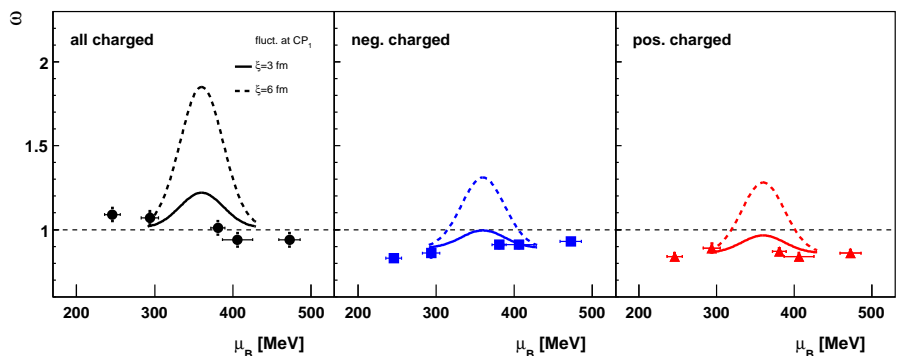}
\vspace{-0.5cm}
\caption[]{Energy dependence of $\omega$ for 1\% most central $Pb+Pb$ collisions
\cite{omega_energy}. Lines correspond to $CP_1$ predictions (see text);
their base-lines are the mean $\omega$ values for 5 energies.}
\label{omegamb}
\end{figure}

\begin{figure}[ht]
\centering
\vspace{-0.3cm}
\includegraphics[scale=0.6]{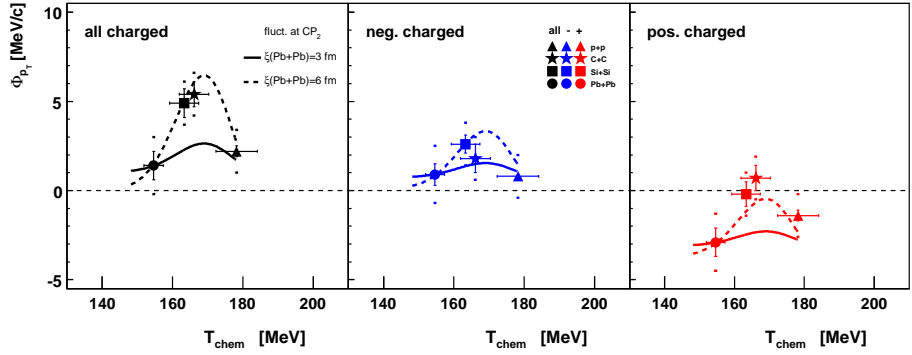}
\vspace{-0.5cm}
\caption[]{System size dependence of $\Phi_{p_T}$ at 158$A$ GeV with 
$p+p$, semi-central $C+C$ (15.3\%) and $Si+Si$ (12.2\%), 5\% most 
central $Pb+Pb$ \cite{fluct_size}. Lines 
correspond to $CP_2$ predictions (see text) shifted to reproduce the 
$\Phi_{p_T}$ value for central $Pb+Pb$ collisions.}
\label{fiptT}
\end{figure}

\begin{figure}[ht]
\centering
\vspace{-0.4cm}
\includegraphics[scale=0.6]{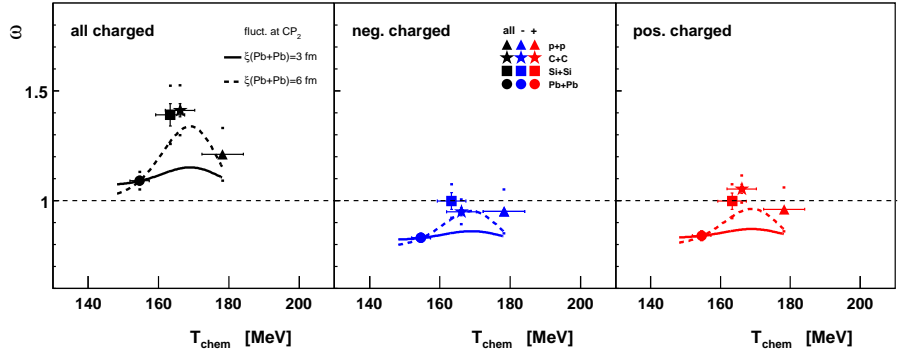}
\vspace{-0.5cm}
\caption[]{System size dependence of $\omega$ at 158$A$ GeV for 1\% most
central $p+p$ \cite{omega_size}, $C+C$ and $Si+Si$ \cite{benjaminPhD}, 
and $Pb+Pb$ \cite{omega_energy}. Lines correspond to $CP_2$ predictions (see
text) shifted to reproduce the $\omega$ value for central $Pb+Pb$ 
collisions. }
\label{omegaT}
\end{figure}

Figures \ref{fiptmb}, \ref{omegamb}, \ref{fiptT}, and \ref{omegaT}
present energy ($\mu_B$) dependence and system size
($T_{chem}$) dependence of $\Phi_{p_T}$ and $\omega$
\footnote{All $\Phi_{p_T}$ and $\omega$ values presented here are 
obtained in the forward-rapidity region and in a limited azimuthal angle 
acceptance (see corresponding papers for details).}. The chemical
freeze-out parameters, $T_{chem}(A,\sqrt{s_{NN}})$ and
$\mu_B(A,\sqrt{s_{NN}})$ were taken from fits with the hadron gas
model \cite{beccatini} to particle yields. The lines correspond to 
predictions for $CP_{1}$ and $CP_{2}$ (Fig.~\ref{cp1cp2}) with estimated 
magnitude of the effects for $\Phi_{p_T}$ and $\omega$ at $CP_{1}$ and 
$CP_{2}$ taken from  Ref.~\cite{SRS, MS} assuming correlation lengths 
$\xi$ decreasing monotonically with decreasing system size: a) 
$\xi$(Pb+Pb) = 6 fm and $\xi$(p+p) = 2 fm (dashed lines) or b) 
$\xi$(Pb+Pb) = 3 fm and $\xi$(p+p) = 1 fm (solid lines). The expected 
magnitudes include NA49 corrections due to limited rapidity range 
(forward-rapidity) and limited azimuthal angle acceptance. The width of 
the enhancement in the ($T, \mu_B$) plane due to the CP is based on 
Ref.~\cite{hatta} and taken as $\sigma (\mu_B) \approx 30$ MeV and 
$\sigma (T) \approx 10$ MeV.

Figures \ref{fiptmb} and \ref{omegamb} show no significant energy
dependence of $p_T$ and multiplicity fluctuations at SPS energies. Thus
the results do not provide evidence for critical point fluctuations, but a
narrower $\mu_B$ scan would be desirable. Figures \ref{fiptT} and
\ref{omegaT} present the system size dependence and 
show a maximum of $\Phi_{p_T}$ and $\omega$ for $C+C$ 
and $Si+Si$ interactions at the top SPS energy. The peak is two 
times higher for all charged than for negatively charged particles as
expected for the critical point \cite{SRS}. Both figures suggest 
that the NA49 data are consistent with the $CP_2$ predictions.

\begin{figure}[ht]
\centering
%\vspace{-0.3cm}
\vspace{-0.4cm}
\includegraphics[scale=0.6]{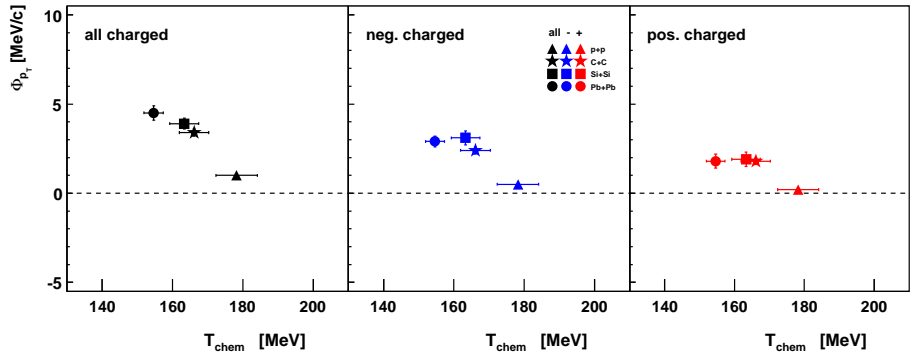}
\vspace{-0.5cm} 
\caption[]{The same as Fig. \ref{fiptT} but low $p_T$ region shown
($0.005 < p_T < 0.5$ GeV/c).}
\vspace{-0.1cm}
\label{low_pt}
\end{figure}

It was expected that fluctuations due to the CP originate mainly 
from low $p_T$ pions \cite{SRS}. Therefore, in the NA49 analysis of 
$\Phi_{p_T}$ the standard $p_T$ range ($0.005 < p_T < 1.5$ GeV/c) was 
divided into two separate $p_T$ regions: $0.5 < p_T < 1.5$ GeV/c and 
$0.005 < p_T < 0.5$ GeV/c. 
Indeed, the high $p_T$ region shows fluctuations consistent with zero 
(see conference slides or Ref.~\cite{kg_hep09}) and  
correlations are observed predominantly at low $p_T$ 
(Fig.~\ref{low_pt}). However, in the low $p_T$ 
region, data do not show a maximum, but a continuous
rise towards $Pb+Pb$ collisions. The origin of this behavior is 
currently being analyzed (short range correlations are considered).

\section{Summary of NA49 results and NA61/SHINE project}

The NA49 experiment obtained  numerous interesting results 
related to both the onset of deconfinement and to the critical 
point. Indications for the onset of deconfinement are seen in 
the energy dependence of the pion yield per wounded nucleon 
$N_W$ (kink), of the $\langle K^{+} \rangle / \langle \pi^{+} \rangle $ 
ratio and $E_s$ (horn), and of the mean transverse mass or inverse 
slope parameters of $m_T$ spectra (step) in central $Pb+Pb$ collisions 
at lower SPS energies (30$A$ GeV). The results are not reproduced by 
hadron-string models (RQMD, UrQMD, HSD). Extension of the HGM fits the 
trend of the $\langle K^{+} \rangle / \langle \pi^{+} \rangle$ ratio but 
relies on unmeasured hadronic states and 
educated-guess assumptions on the branching ratios. Also the energy 
dependence of event-by-event hadron ratio fluctuations shows interesting 
effects in the lower SPS energy range (increase of $\sigma_{dyn}(K/\pi)$ 
and sign change of $\sigma_{dyn}(K/p)$), but their relation to the onset 
of deconfinement is not clear.
Other (not shown) observables were examined for signatures
of the onset of deconfinement: a minimum of sound velocity at 
low SPS energies as expected for a 1$^{st}$ order phase transition 
\cite{volker_cpod09}, azimuthal correlations and disappearance of 
near-side correlations~\cite{szuba_qm09}. An additional energy and 
system-size scan is important to search for the onset of deconfinement 
in collisions of light nuclei. This is the purpose of the 
NA61/SHINE experiment \cite{shine} at the CERN SPS.

NA49 also searched  for indications of the critical point. There are 
no indications of the CP in the energy dependence of multiplicity and mean 
$p_T$ fluctuations in central $Pb+Pb$ collisions.
Other (not shown) observables were studied: ratios of the 
anti-baryon/baryon transverse mass spectra and elliptic flow $v_2$.
Neither shows indications of CP \cite{kg_qm09}. 
However, the system size dependence of mean $p_T$ and multiplicity fluctuations 
at 158$A$ GeV shows a maximum in the complete $p_T$ range 
(consistent with $CP_2$ predictions) and an increase from $p+p$ up to 
$Pb+Pb$ collisions in the low $p_T$ region. The low 
$p_T$ region will be carefully analyzed for the effects of short range 
correlations on $\Phi_{p_T}$ and $\omega$. A detailed energy and 
system-size scan is necessary to establish the existence of the 
critical point. Therefore, the CP search will be continued by the 
NA61/SHINE experiment.

\begin{figure}[ht]
\centering
\vspace{-0.2cm}
\includegraphics[scale=0.8]{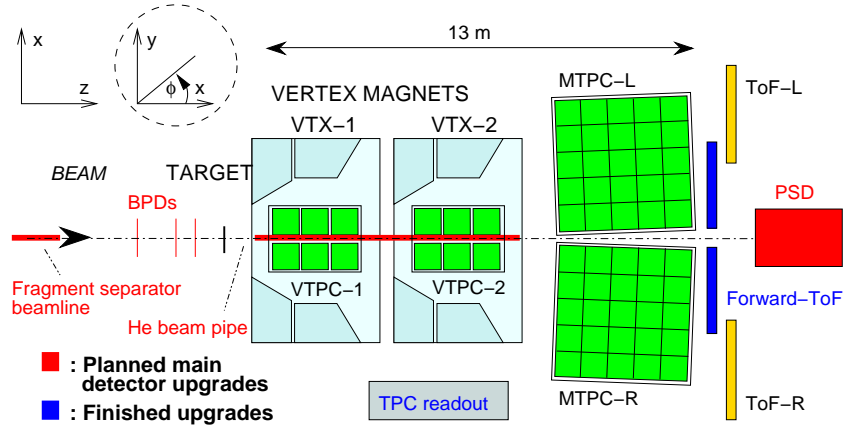}
\caption[]{The NA61/SHINE setup \cite{shine}.}
\label{na61setup}
\end{figure}

The NA61/SHINE \footnote{SHINE $-$ SPS Heavy Ion and Neutrino Experiment} 
experiment (Fig.~\ref{na61setup}) is the successor of NA49. The main 
detector components are inherited from NA49. Several  upgrades 
were already completed or are planned in the future: a) in 2007 a forward $ToF$ wall 
was constructed to extend $ToF$ acceptance for particles with $p<3$ 
GeV/c, b) in 2008 the TPC read-out and Data Aqusition system were 
upgraded to increase the event recording rate by a factor of $\approx$10, c) in 
2011 the NA49 Forward Calorimeter will be replaced by the Projectile 
Spectator Detector (PSD) with a five times better resolution.
This excellent resolution (about one nucleon in the studied energy 
range) is crucial especially for the analysis of multiplicity 
fluctuations.

In the NA61/SHINE experiment hadron production in $p+p$, $p+A$, $h+A$, and 
$A+A$ reactions at various energies will be analyzed. A broad experimental 
program is planned: search for the critical point, study of the 
properties of the onset of deconfinement, high $p_T$ physics (energy 
dependence of the nuclear modification factor), and analysis of hadron 
spectra for the T2K neutrino experiment and for the Pierre Auger 
Observatory and KASCADE cosmic-ray experiments. Within the NA61/SHINE ion 
program we plan, for the first time in history, to perform a 2D scan 
with system size and energy. The future data on $p+p$ 
(2009-2010; data taking in progress), $^{11}B+^{12}C$ (2010-2013), 
$^{40}Ar+^{40}Ca$ (2012), and $^{129}Xe+^{139}La$ (2014) 
will allow to cover a broad range of the phase diagram (see 
Fig.~\ref{cp1cp2}). With these data we will be able to:

\begin{enumerate}
\item Study the properties of the onset of deconfinement. In principle 
one can search for the onset of the 'horn', 'kink', 'step' in collisions 
of light nuclei (the structures observed for $Pb+Pb$/$Au+Au$ should 
vanish with decreasing system size)  
\item Search for the critical point. An increase of the critical point signal, 
the so-called hill of fluctuations, is expected for systems freezing-out near 
the critical point. Therefore non-monotonic dependence of the CP signal on 
control parameters (energy, centrality, ion size) can help to locate 
the critical point.  
\end{enumerate}

The NA61 program will be complemented by the efforts of other 
international and national laboratories, BNL RHIC 
($ 5 < \sqrt{s_{NN}} < $ 39 GeV), JINR NICA ($ 3 < \sqrt{s_{NN}} < $ 9 GeV), 
GSI SIS-100(300) ($ 2.3 < \sqrt{s_{NN}} < $ 8.5 GeV) and by the heavy 
ion program at the CERN LHC ($\sqrt{s_{NN}}$ = 5500 GeV and 
$\sqrt{s}$ = 14000 GeV for $p+p$).

%%%%%%%%%%%%%%%%%%%%%%%%%%%%%%%%%%%%%%%%%%%%%%%%%

\vspace{1cm}

\noindent
{\bf Acknowledgements (NA49):} This work was supported by
the US Department of Energy Grant DE-FG03-97ER41020/A000,
the Bundesministerium fur Bildung und Forschung, Germany, 
the Virtual Institute VI-146 of \\ Helmholtz Gemeinschaft, Germany,
the Polish Ministry of Science and Higher Education (1 P03B 006 30, 1 
P03B 127 30, 0297/B/H03/2007/33, N N202 078735),
the Hungarian Scientific Research Foundation (T032648, T032293, 
T043514),
the Hungarian National Science Foundation, OTKA, (F034707),
the Korea Research Foundation (KRF-2007-313-C00175),
the Bulgarian National Science Fund (Ph-09/05),
the Croatian Ministry of Science, Education and Sport (Project 
098-0982887-2878)
and
Stichting FOM, the Netherlands.

\vspace{0.5cm}

\noindent
{\bf Acknowledgements (NA61/SHINE):} This work was supported by  
the Hungarian Scientific Research Fund (OTKA 68506),
the Polish Ministry of Science and Higher Education (N N202 3956 33),
the Federal Agency of Education of the Ministry of Education and Science
of the Russian Federation (grant RNP 2.2.2.2.1547) and
the Russian Foundation for Basic Research (grants 08-02-00018 and 
09-02-00664),
the Ministry of Education, Culture, Sports, Science and Technology,
Japan, Grant-in-Aid for Scientific Research (18071005, 19034011,
19740162),
Swiss Nationalfonds Foundation 200020-117913/1 
and ETH Research Grant TH-01 07-3.

\vspace{0.5cm} 
\noindent
Katarzyna Grebieszkow would like to thank the organizers 
of XXXI Mazurian Conference on Physics for inviting her and giving the 
opportunity to present the NA49 results and the NA61/SHINE plans.


\clearpage

\begin{thebibliography}{00}

\bibitem{beccatini} F. Beccatini, J. Manninen, M. Gazdzicki, {\it
Phys. Rev.} {\bf C73} (2006) 044905.

\bibitem{fodor_latt_2004} Z. Fodor and S. D. Katz, {\it JHEP} {\bf
0404} (2004) 050.

\bibitem{na49_nim} S. Afanasiev et al. (NA49 Collab.), {\it Nucl.
Instrum. Meth.} {\bf A430} (1999) 210.

\bibitem{mg_model} M. Gaździcki, M. Gorenstein, {\it Acta Phys. Polon.} 
{\bf B30} (1999) 2705. 

\bibitem{kpi_paper} C. Alt et al. (NA49 Collab.), {\it Phys. Rev.} 
{\bf C77} (2008) 024903.

\bibitem{star_lowen} L. Kumar et al. (STAR Collab.), arXiv:0812.4099. 
%(SQM2008); Kumar (QM09) 
%arXiv:0907.4504 (plenary QM09) 

\bibitem{HGM_1} A. Andronic, P. Braun-Munzinger, J. Stachel, {\it Nucl. 
Phys.} {\bf A772} (2006) 167.

\bibitem{HGM_2} A. Andronic, P. Braun-Munzinger, J. Stachel, 
{\it Acta Phys. Polon.} {\bf B40} (2009) 1005.

\bibitem{PDG2008} C. Amsler et al. (Particle Data Group), {\it Phys. 
Lett.} {\bf B667} (2008) 1. 

\bibitem{smes_temp} M. Gorenstein et al., {\it Phys. Lett.} {\bf B567} 
(2003) 175. 

\bibitem{smes_hydro} Y. Hama et al., {\it Braz. J. Phys.} {\bf 34} 
(2004) 322. 

\bibitem{kpi_fluct} % published:  arXiv:0808.1237,
T. Schuster et al. (NA49 Collab.), {\it} PoS {\bf CPOD2009} 
(2009) 029, and references therein.


\bibitem{SRS} M. Stephanov, K. Rajagopal, E. V. Shuryak,
{\it Phys. Rev.} {\bf D60} (1999) 114028.


\bibitem{kg_qm09} K. Grebieszkow et al. (NA49 Collab.), arXiv:0907.4101, 
and references therein.

\bibitem{fluct_size} T. Anticic et al. (NA49 Collab.), {\it Phys. Rev.}
{\bf C70} (2004) 034902.

\bibitem{fluct_energy} T. Anticic et al. (NA49 Collab.), {\it Phys.
Rev.} {\bf C79} (2009) 044904.

\bibitem{omega_size} C. Alt et al. (NA49 Collab.), {\it Phys. Rev.}
{\bf C75} (2007) 064904.

\bibitem{omega_energy} C. Alt et al. (NA49 Collab.), {\it Phys. Rev.}
{\bf C78} (2008), 034914.

\bibitem{konchak} V. Konchakovski et al., {\it Phys. Rev.} {\bf C73}
(2006) 034902, and priv. comm.

\bibitem{benjaminPhD} B. Lungwitz, PhD thesis (2008),
https://edms.cern.ch/document/989055/1 
%(unpublished).

\bibitem{MS} M. Stephanov, private communication.

\bibitem{hatta} Y. Hatta, T. Ikeda, {\it Phys. Rev.} {\bf D67} (2003)
014028.

\bibitem{kg_hep09} K. Grebieszkow et al. (NA49 and NA61 Collab.),   
arXiv0909.0485.

\bibitem{volker_cpod09} V. Friese et al. (NA49 Collab.), 
arXiv:0908.2720.

\bibitem{szuba_qm09} M. Szuba et al. (NA49 Collab.), arXiv:0907.4403.

\bibitem{shine} https://na61.web.cern.ch/na61/xc/index.html


\end{thebibliography}
\end{document}